\documentclass[11pt]{article}
\usepackage[dvips]{epsfig}
\usepackage[T1]{fontenc}
\usepackage[latin1]{inputenc}
\usepackage{graphicx}
\usepackage[english]{babel}
\usepackage{amsmath}
\usepackage{amssymb}
\usepackage{amsfonts}
\usepackage[T1]{fontenc}
\usepackage{color}
\usepackage{babel}
\usepackage{verbatim}
\usepackage[unicode=true,pdfusetitle,bookmarks=true,bookmarksnumbered=false,bookmarksopen=false,
 breaklinks=false,pdfborder={0 0 1},backref=false,colorlinks=true]{hyperref}
\hypersetup{linkcolor=blue,citecolor=blue}
\makeatletter
\usepackage[dvips]{epsfig}
\usepackage[T1]{fontenc}
\usepackage{color}
\usepackage{pdflscape}
\usepackage{cite}

\title{\bf Signature Change in $f(R, T_\phi)$ Theory}

\author{Serkan Doruk Hazinedar$^a$\thanks{
email: doruk.hazinedar@bilkent.edu.tr }, ~Yaghoub Heydarzade$^a$\thanks{
email: yheydarzade@bilkent.edu.tr}, \\
\\{\small $^a$Department of Mathematics, Faculty of Sciences, Bilkent University, 06800 Ankara, T{\" u}rkiye}}

\date{}

\begin{document}

\maketitle

\begin{abstract}
We investigate a simple $f(R, T_\phi)$ gravity model coupled to a scalar field and demonstrate that the theory admits classical degenerate metric solutions, analogous to those known in general relativity. In particular, we identify a class of solutions that exhibits a smooth transition from a Euclidean to a Lorentzian domain, thus yielding a classical dynamical realization of signature change.
\\
\\
\textit{Dedicated to Metin G\" urses on the occasion of his becoming an emeritus professor after 60 years of research and teaching.}
\end{abstract}

\section{Introduction}

Einstein's theory of gravitation is traditionally formulated on a smooth four-dimensional manifold endowed with a Lorentzian metric structure. The causal structure implied by Lorentzian signature plays a central role in both classical and quantum field theory. However, when gravitational dynamics are examined in the context of quantum cosmology, the pre-eminence of Lorentzian geometry becomes less absolute. In canonical quantum gravity, as originally formulated by DeWitt \cite{DeWitt1967}, the configuration space of three-geometries (superspace) naturally accommodates histories that need not correspond globally to Lorentzian spacetimes. Similarly, early path-integral approaches to quantum cosmology, including Misner's sum over histories framework \cite{Misner1969}, suggested that semiclassical states may arise from superpositions of geometries with differing topological or metric properties. In particular, the Hartle-Hawking \textit{no-boundary proposal} \cite{HartleHawking1983} advocates that the universe may be described semiclassically by compact Riemannian geometries that smoothly match onto Lorentzian spacetime. These considerations motivate the study of spacetimes admitting metric signature change, where the metric becomes degenerate on a hypersurface separating Euclidean and Lorentzian regions.

Early investigations of signature change focused on quantum cosmological path-integral constructions \cite{HartleHawking1983}. Subsequently, classical realizations of signature change were explored within general relativity by allowing degenerate metrics as solutions of Einstein's field equations. In a seminal work, Dereli and Tucker \cite{Dereli} demonstrated that Einstein gravity coupled to a self-interacting scalar field admits analytic solutions in which the metric smoothly undergoes a transition from Euclidean to Lorentzian signature across a hypersurface where the determinant of the metric vanishes linearly. Within canonical quantum cosmology, this theme was sharpened by the analysis of Wheeler--DeWitt wave packets in minisuperspace, where specific quantum states were shown to peak on classical branches admitting Euclidean--Lorentzian transition \cite{DereliOnderTucker}, and later by perfect-fluid FLRW models in which Schutz time leads to a Schr\"odinger--Wheeler--DeWitt description with explicit signature-changing expectation values for the scale factor \cite{PedramJalalzadeh}.

The geometric structure of such transitions was later clarified in the framework developed by Kossowski and Kriele \cite{Kriele}, who provided precise regularity conditions for transverse signature-changing spacetimes. In this formulation, the determinant of the metric vanishes to first order on the transition hypersurface, the radical direction is transverse, and the hypersurface is totally geodesic, thereby preventing the appearance of distributional curvature sources. These geometric conditions ensure that the transition is smooth in a well-defined differential sense.

Beyond the Einstein-scalar theory, signature change has been examined in several extended gravitational frameworks. In higher-dimensional and Kaluza--Klein cosmologies, classical Euclidean-Lorentzian transitions arise naturally within constrained minisuperspace models, sometimes in connection with compactification mechanisms \cite{DarabiSepangi1999}. In Einstein-Cartan theory, torsion contributions can modify the early universe dynamics and permit non-standard phase structures including Euclidean regimes \cite{Gasperini1986, Randono2014}. In \cite{VakiliJalalzadeh2013} it was shown that in an Einstein--Cartan FLRW model with Weyssenhoff perfect fluid that spin--spin interactions can generate smooth classical solutions with a degenerate metric, describing a continuous transition from a finite Euclidean region to a Lorentzian domain. Furthermore, effective approaches inspired by loop quantum cosmology indicate that quantum-geometry corrections may induce a transition between hyperbolic and elliptic regimes of the effective field equations, interpretable as a change of signature at high curvature scales \cite{BojowaldPaily2012,Mielczarek2014}. These results suggest that signature change is not peculiar to Einstein gravity with scalar field source, but may arise more generally in modified or extended gravitational theories.

In parallel with these developments, modified gravity theories have been extensively studied as potential explanations for late-time cosmic acceleration, early-universe inflation, and deviations from general relativity at large scales. Among these models, $f(R,T)$ gravity, proposed by Harko \emph{et al.} \cite{Harko}, introduces an explicit dependence of the gravitational action on the trace $T$ of the matter energy-momentum tensor. The trace coupling induces a direct interaction between matter and geometry, modifying the effective gravitational field equations and leading to non-standard cosmological dynamics even in spatially homogeneous and isotropic settings.

Despite the extensive literature on $f(R,T)$ cosmology, the question of whether classical signature change can occur in such trace-coupled theories has not been analyzed in the literature, and this gap constitutes the main novelty of the present work. Since the trace coupling alters both the effective stress-energy tensor and the scalar-field dynamics, it is not \emph{a priori} clear whether the regular signature-changing branch found in the Einstein-scalar case survives, nor how the Kossowski--Kriele regularity conditions are modified in the presence of matter-geometry coupling.

In this work, we investigate signature change in a minimally coupled $f(R,T_\phi)$ model of the form $
f(R,T_\phi) = R + \alpha T_\phi$
where $T_\phi$ is the trace of the scalar field energy-momentum tensor. Our approach closely follows the analytic approach of Dereli and Tucker \cite{Dereli}. We employ a signature-evolution parameter $\beta$ to describe a type-changing FLRW geometry and seek real solutions that extend smoothly across the hypersurface $\Sigma_0$ ($\beta=0$). After performing a minisuperspace reduction, the system can be mapped via a hyperbolic field redefinition to an effective oscillator-type model. The Hamiltonian constraint inherited from time-reparameterization invariance then selects the physically admissible branches, analogously to the ``zero-energy'' sector in \cite{Dereli}. However, the trace-coupling parameter $\alpha$ introduces qualitatively new dynamical regimes absent in general relativity. In particular, special values of $\alpha$ correspond to degenerate scalar limits or kinetic-dominated sectors, modifying both the effective potential and the minisuperspace structure. We show that regular transverse signature change remains possible for suitable parameter choices, and that the resulting solutions satisfy the Kossowski--Kriele regularity conditions \cite{Kriele}, ensuring that no distributional source terms arise at the transition hypersurface.

In addition to quadratic type potentials analogous to the Sinh-Gordon structure of \cite{Dereli}, we analyze an exponential potential motivated by higher-curvature and higher-dimensional theories \cite{LucchinMatarrese1985, Halliwell1987, Wetterich1988, Copeland1998}. For this class, we obtain exact solutions and prove a no-go theorem excluding non-singular Lorentzian bounces within the signature-changing branch. Thus, while the model admits smooth Euclidean-Lorentzian transitions, it does not mediate a contracting to expanding bounce in the Lorentzian domain.

Our results demonstrate that classical signature change persists in trace-coupled modified gravity, but with a modified branch structure and additional constraints arising from matter-geometry coupling. The $f(R,T_\phi)$ framework therefore provides a natural and analytically tractable extension of signature dynamics beyond general relativity.

\section{ $f(R,T_\phi)$ Model}

Consider the signature-changing FLRW line element
\begin{equation}\label{metric1}
ds^2 =-\beta\,d\beta^2+\tilde a(\beta)^2\,\gamma_{ij}dx^i dx^j= -\beta \, d\beta^2 + \tilde{a}^2(\beta) \left( 
\frac{dr^2}{1 - k r^2} 
+ r^2 d\Omega^2 
\right),
\end{equation}
where the spatial curvature $k = 0, \pm 1$ represents a flat, closed, or open universe, respectively. 
The signature of the metric is Lorentzian for $\beta > 0$ and Euclidean for $\beta < 0$. For the Lorentzian region, one can recover the cosmic time $t$ by
\[
t = \frac{2}{3} \beta^{3/2},
\]
leading to
\begin{equation}\label{metric2}
ds^2 = -dt^2 + a^2(t)\left( 
\frac{dr^2}{1 - k r^2} 
+ r^2 d\Omega^2 
\right),
\end{equation}
where $a(t) = \tilde{a}(\beta(t))$. Following \cite{Dereli}, it is convenient to formulate our differential equations in a region that does not include $\beta = 0$ and look for real solutions for $a(t)$ smoothly passing through the $\beta = 0$ hypersurface $\Sigma_0=\Sigma_\beta|_{\beta=0}$. The curvature scalar corresponding to metric \eqref{metric2} reads
\begin{equation}
R=
6\left( \frac{\ddot{a}}{a}
+
\frac{k + \dot{a}^2}{a^2}\right).
\end{equation}

We consider a minimally coupled $f(R,T_\phi)$ action with a scalar field source (with $\kappa=1$)\cite{Harko}
\begin{eqnarray}\label{action}
S&=&\int d^4x \sqrt{-g} \left(f(R, T_\phi)\right)+\int d^4x \sqrt{-g} L_m\nonumber\\
&=&\int d^4x \sqrt{-g}
\left(
\frac{1}{2}R
+ f(T_\phi)\right)+
\int d^4x \sqrt{-g}
\left(-\frac{1}{2}\omega(\phi)\partial_\mu\phi\partial^\mu\phi
- V(\phi)
\right).
\end{eqnarray}
In the present work, we shall analyze the simple linear case $f(T_\phi)=\alpha T_\phi$. 
Considering the homogeneous and isotropic FLRW metric,
the scalar field will be only time dependent, i.e., $\phi=\phi(t)$, and hence the trace of energy-momentum source will be 
\begin{equation}
T_\phi=-\omega(\phi)\dot \phi^2 -4V(\phi).
 \end{equation}
 
Variation with respect to the metric yields the field equations 
\begin{eqnarray}
    G_{\mu \nu}=T_{\mu \nu}^{\text{eff}},
\end{eqnarray}
where $G_{\mu \nu}$ is the Einstein tensor and the effective energy-momentum tensor reads
\begin{equation}
T^{\mathrm{eff}}_{\mu\nu}
=
(1+2\alpha)\,\omega(\phi)\,\partial_\mu\phi\,\partial_\nu\phi
-
g_{\mu\nu}\left[
\left(\frac{1}{2}+\alpha\right)\omega(\phi)\,\partial_\rho\phi\,\partial^\rho\phi
+
(1+4\alpha)\,V(\phi)
\right].
\end{equation}
Under the ADM decomposition, the totally geodesic condition on the signature-change hypersurface $\Sigma_0$ is equivalent to the vanishing of the extrinsic curvature, \(K_{ij}\big|_{\Sigma_0}=0\), which in turn implies $\partial_t h_{ij}\big|_{\Sigma_0}=0$ \cite{Kriele}. In FLRW, where $h_{ij}=a^2(t)\gamma_{ij}$, this reduces to $\dot a\big|_{\Sigma_0}=0$.
Then, under the Kossowski--Kriele assumptions for a transverse type-changing $C^k$ spacetime \cite{Kriele}, when $k \geq 3$ boundedness of the energy-momentum tensor is naturally interpreted as boundedness of $T_{\mu \nu}^{\text{eff}}$. Hence, the geometric conditions remain the controlling conditions for regularity across the signature-change hypersurface $\Sigma_0$, now in the presence of the $\alpha T_\phi$ coupling. 
Meanwhile, variation with respect to $\phi$ yields the effective Klein--Gordon equation for a non-canonical scalar field
\begin{equation}
(1+2\alpha)\,\omega(\phi)\,\Box\phi
+\left(\frac{1}{2}+\alpha\right)\omega_{,\phi}(\phi)\,\partial_\mu\phi\partial^\mu\phi
-(1+4\alpha)V_{,\phi}(\phi)=0.
\end{equation}
Two special values of the linear trace-coupling parameter, $\alpha=-\tfrac12$ and $\alpha=-\tfrac14$, lead to qualitatively distinct regimes. For $\alpha=-\tfrac12$, the coefficient of the kinetic term in the scalar-field equation vanishes, so the effective Klein--Gordon equation ceases to be an evolution equation and reduces instead to the constraint $V_{,\phi}=0$. In this degenerate limit, the effective energy--momentum tensor collapses to the vacuum-like form $T^{\mathrm{eff}}_{\mu\nu}=V(\phi)\,g_{\mu\nu}$, i.e. an effective cosmological-constant source. By contrast, for $\alpha=-\tfrac14$, the potential decouples from the effective field equation, so the scalar sector remains dynamical but is driven purely by its kinetic/non-canonical structure. In this sense, $\alpha=-\tfrac12$ defines a degenerate (non-propagating) scalar limit, whereas $\alpha=-\tfrac14$ defines a kinetic-dominated critical regime.

From the total Lagrangian in \eqref{action}, one finds the minisuperspace Lagrangian as
\begin{equation}\label{lag1}
L(a, \dot a , \phi, \dot \phi)=
-3a\dot a^2
+3ka
-A a^3\omega(\phi)\dot\phi^2
-B a^3 V(\phi),
\end{equation}
where $A=\alpha+\frac{1}{2}$ and $B=4\alpha +1$ are coupling constants from the theory.
The underlying mechanical behavior can be better understood by the hyperbolic field redefinition \cite{Dereli}
\begin{equation}
X=a^{3/2}\cosh(\sigma\phi),
\qquad
Y=a^{3/2}\sinh(\sigma\phi).
\end{equation}
Applying these transformations, the Lagrangian takes the form
\begin{eqnarray}\label{lag2*}
L(X,Y,\dot X,\dot Y)
&=&
-\frac{4}{3}\,
\frac{(X\dot X - Y\dot Y)^2}{X^2 - Y^2}
-\frac{A}{\sigma^2}\,\omega\!\big(\phi(X,Y)\big)
\frac{(X\dot Y - Y\dot X)^2}{X^2 - Y^2}\nonumber\\
&&+3k\,(X^2 - Y^2)^{1/3}
- B\,(X^2 - Y^2)\,V\!\big(\phi(X,Y)\big).
\end{eqnarray}
Here, the two kinetic terms in \eqref{lag1} and \eqref{lag2*} can be combined and simplified to 
\begin{equation}
-3a\dot a^2 - A a^3\omega\dot\phi^2
=
-\frac{4}{3}(\dot X^2 - \dot Y^2).
\end{equation}
provided
\begin{equation}
\sigma^2 = -\frac{3}{4}A\omega
\quad \text{ and ~~$\omega =constant.$ }
\end{equation}
The hyperbolic map demands the real values of $\sigma$, hence the coupling of the theory must obey the constraint $A\omega<0$. 
Hence, the Lagrangian reduces to the form
\begin{equation}\label{lag2}
\mathcal{L}(X,Y,\dot X,\dot Y)
=
\dot X^2 - \dot Y^2
-\frac{9k}{4}(X^2-Y^2)^{1/3}
+\frac{3B}{4}(X^2-Y^2)V(\phi(X, Y)).
\end{equation}
where we defined $\mathcal{L}:=-\frac34 L$. Following \cite{Dereli}, we consider the spatially flat universe $(k=0)$, and select the potential that satisfies
\begin{equation}
\frac{3B}{4}(X^2-Y^2)V(\phi)
=
\beta_1 X^2+\beta_2 Y^2+2\beta_3 XY,
\end{equation}
where $\beta_1, \beta_2$ and $\beta_3$ are constant parameters.
Hence, the effective Lagrangian simplifies to
\begin{equation}
\mathcal{L}(X,Y,\dot X,\dot Y)
=
\dot X^2 - \dot Y^2
+\beta_1 X^2+\beta_2 Y^2+2\beta_3 XY.
\end{equation}
This Lagrangian can also be time-reparametrized with $\tau$ and an embedding function $t=N(\tau)$ and 
$dt = N'(\tau)\, d\tau$, as
\begin{equation}
\mathcal{L}_\tau(X,Y,\dot X,\dot Y, N')
=
N' \left[\left(\frac{X'}{N'}\right)^2 - \left(\frac{Y'}{N'}\right)^2 + \beta_1 X^2+\beta_2 Y^2+2\beta_3 XY\right].
\label{eq:L_tau}
\end{equation}
The Euler--Lagrange equations for $X,Y$ and $N$ are
\begin{eqnarray}
X''-\frac{N''}{N'}\,X'&=&(N')^2(\beta_1 X+\beta_3 Y),
\label{eq:X_eq_expanded}\nonumber\\
Y''-\frac{N''}{N'}\,Y'&=&-(N')^2(\beta_2 Y+\beta_3 X),\nonumber\\
-\Big(\frac{X'^2-Y'^2}{(N')^2}\Big)+\beta_1 X^2+\beta_2 Y^2+2\beta_3 XY&=&C.
\label{eq:Y_eq_expanded}
\end{eqnarray}
The third equation is indeed the Hamiltonian constraint equation and $C$ is an integration constant. 
The above system of differential equations can be written in a compact matrix form as
\begin{equation}\label{ode}
\psi'' = (N')^2 M \psi + \frac{N''}{N'} \psi',
\end{equation}
where 
\[
\psi=
\begin{pmatrix}
X\\
Y
\end{pmatrix},
\quad
M=
\begin{pmatrix}
\beta_1 & \beta_3\\
-\beta_3 & -\beta_2
\end{pmatrix}.
\]
Here,  one observes that choosing a convenient gauge $N'=1$ removes the friction-like term $\frac{N''}{N'} \psi'$, while the Hamiltonian constraint sets C = 0. Moreover, since the minisuperspace model is inherited from a time-reparameterization invariant gravitational theory \cite{Harko}, the lapse equation imposes the Hamiltonian constraint.
Thus, the integration constant in \eqref{eq:Y_eq_expanded} must be set to \(C=0\), corresponding to the "zero-energy set" in the terminology of \cite{Dereli}, and the system reduces to a linear second-order matrix ODE with constant coefficients.


Before discussing the solutions, let us analyze the potential. Substituting the transformation relations for $X$ and $Y$, the potential reads as
\begin{equation}
V(\phi)
=
\frac{4}{3B}
\left[
\beta_1
+(\beta_1+\beta_2)\sinh^2(\sigma\phi)
+\beta_3\sinh(2\sigma\phi)
\right].
\end{equation}
The constant term provides a vacuum-energy offset \(V(0)=\frac{4\beta_1}{3B}\). For \(\beta_3=0\), the potential is even in \(\phi\) and \(\phi=0\) is an extremum; in that case the effective mass is  
\[
m^2=\left.V''(\phi)\right|_{\phi=0}=\frac{8\sigma^2}{3B}(\beta_1+\beta_2).
\]
Here, one notes that due to the coupling $B$ the mass is rescaled in comparison to GR \cite{Dereli}.  For \(\beta_3\neq 0\), the \(\beta_3\)-term generates a tadpole, breaks \(\phi\to-\phi\), and shifts the extremum away from \(\phi=0\); the physical mass should then be evaluated at the shifted vacuum.
The extremum condition $
\left.\frac{dV}{d\phi}\right|_{\phi_*}=0
$
reduces to
\[
\tanh(2\sigma\phi_*)
=
-\frac{2\beta_3}{\beta_1+\beta_2}.
\]
Hence, we must require $\beta_1+\beta_2\neq 0$ so that a nontrivial stationary point can exist. 
Furthermore, since $|\tanh(2\sigma\phi_*)|<1$ for real $\phi_*$, reality of the extremum demands
\[
\left|\frac{2\beta_3}{\beta_1+\beta_2}\right|<1
\quad\Longleftrightarrow\quad
(\beta_1+\beta_2)^2>4\beta_3^2,
\]
otherwise no finite real stationary point exists and the potential exhibits runaway behavior. Under these conditions, the exact location of the extremum is
\[
\phi_*=
\frac{1}{2\sigma}
\operatorname{arctanh}\!\left(
-\frac{2\beta_3}{\beta_1+\beta_2}
\right).
\]
The value of the potential at this point is
\[
V(\phi_*)
=
\frac{4}{3B}
\left[
\frac{\beta_1-\beta_2}{2}
+
\frac{\operatorname{sgn}(\beta_1+\beta_2)}{2}
\sqrt{(\beta_1+\beta_2)^2-4\beta_3^2}
\right].
\]
The extremum is a true minimum of $V$ iff
$\left.\frac{d^2V}{d\phi^2}\right|_{\phi_*}>0,$
which requires
\[
\operatorname{sgn}\!\left(\frac{\sigma^2}{B}\right)
\operatorname{sgn}\!\left(
\frac{(\beta_1+\beta_2)^2-4\beta_3^2}
{\beta_1+\beta_2}
\right)
>0,
\]
together with the existence condition $
(\beta_1+\beta_2)^2>4\beta_3^2$. In the common case $\sigma^2>0$, this simplifies to
\[
\begin{cases}
B>0 \;\Rightarrow\; \beta_1+\beta_2>0,\\[4pt]
B<0 \;\Rightarrow\; \beta_1+\beta_2<0,
\end{cases}
\qquad
\text{with } (\beta_1+\beta_2)^2>4\beta_3^2.
\]
So the coupling $B=4\alpha +1$ included in the $f(R, T_\phi)=R+\alpha T_\phi$ theory sets the overall strength of the potential and plays an important role in the scalar field dynamics.

One can find the eigenvalues of coefficient matrix $M$ determining the dynamics of the system \eqref{ode} as
\[
\lambda_{\pm}
=
\frac{\beta_1 - \beta_2}{2}
\;\pm\;
\frac{1}{2}
\sqrt{(\beta_1 + \beta_2)^2 - 4\beta_3^2}.
\]
These determine whether solutions are exponential, oscillatory,
or mixed hyperbolic--oscillatory.
For each eigenvalue $\lambda_{\pm}$, an eigenvector is found as
\[
v=
\begin{pmatrix}
\beta_3 \\
\lambda_{\pm}-\beta_1
\end{pmatrix}.
\]
Hence the general solution can be written as
\[
\psi(t)
=
\sum_{i=\pm}
\left[
C_i e^{\sqrt{\lambda_i}\,t}
+
D_i e^{-\sqrt{\lambda_i}\,t}
\right]
\begin{pmatrix}
\beta_3 \\
\lambda_i-\beta_1
\end{pmatrix},
\]
where $C_i, \, D_i$ are integration constants. Explicitly,
\begin{eqnarray}
X(t) &=&
\sum_{i=\pm}
\left[
C_i e^{\sqrt{\lambda_i}\,t}
+
D_i e^{-\sqrt{\lambda_i}\,t}
\right]\beta_3,
\nonumber\\
Y(t) &=&
\sum_{i=\pm}
\left[
C_i e^{\sqrt{\lambda_i}\,t}
+
D_i e^{-\sqrt{\lambda_i}\,t}
\right](\lambda_i-\beta_1).
\end{eqnarray}

Considering solutions satisfying $\dot\psi(0)=0$, i.e., $\dot X(0)=\dot Y(0)=0$, one finds $C_i=D_i$ for each $i=\pm$. This implies
\begin{eqnarray}
X(t) &=& 2\beta_3 \sum_{i=\pm} C_i \cosh(\sqrt{\lambda_i}\,t), \nonumber\\
Y(t) &=& 2 \sum_{i=\pm} C_i (\lambda_i-\beta_1)\cosh(\sqrt{\lambda_i}\,t).
\end{eqnarray}
Having obtained the general solutions $X(t),Y(t)$ of the equations of motion, we should now impose the Hamiltonian constraint \eqref{eq:Y_eq_expanded} evaluated at the signature change hypersurface $\Sigma_0$ to restrict the integration constants as follows
\begin{equation*}
\beta_1\beta_3^2\left(\sum_{i=\pm} C_i\right)^2
+\beta_2\left(\sum_{i=\pm} C_i(\lambda_i-\beta_1)\right)^2
+2\beta_3^2\left(\sum_{i=\pm} C_i\right)\left(\sum_{i=\pm} C_i(\lambda_i-\beta_1)\right)=0.
\end{equation*}
Equivalently, when $\beta_1\neq 0$, one can rewrite it as
\begin{equation}
\beta_1\beta_3^2
\left[
\left(\sum_{i=\pm} C_i\right)
-r_+\left(\sum_{i=\pm} C_i(\lambda_i-\beta_1)\right)
\right]
\left[
\left(\sum_{i=\pm} C_i\right)
-r_-\left(\sum_{i=\pm} C_i(\lambda_i-\beta_1)\right)
\right]=0,
\end{equation}
where $r_\pm=\frac{-1\pm\sqrt{1-\beta_1\beta_2/\beta_3^2}}{\beta_1}$, and for real $r_\pm$ we assume $\beta_3^2\geq \beta_1\beta_2$. Then, it provides the following relation
\begin{eqnarray}
    \frac{X(0)}{Y(0)}= \beta_3 \Bigg(\frac{\sum_{i=\pm} C_i}{\sum_{i=\pm} C_i(\lambda_i-\beta_1)}\Bigg)= \beta_3 r_{\pm}.
\end{eqnarray}
Therefore, the Hamiltonian constraint does not merely reduce the number of integration constants; it splits the time-symmetric solutions into two distinct branches. In particular, at the signature change hypersurface $\Sigma_0$, the initial data $(X(0),Y(0))$ must satisfy $X(0)=\beta_3 r_\pm\, Y(0)$, so the admissible initial points lie on one of two straight lines in the \((X,Y)\)-plane. Equivalently, for each branch the ratio \(C_-/C_+\) is fixed algebraically, yielding two classes of regular solutions.

Finally, the functions $a(t)$ and $\phi(t)$ can be recovered from $X(t)$ and $Y(t)$ as
\begin{eqnarray*}
    a(t)&=&\Big(X^2(t)-Y^2(t)\Big)^\frac{1}{3},\\
    \phi(t)&=&\frac{1}{\sigma}\operatorname{arctanh}\left(\frac{Y(t)}{X(t)}\right).
\end{eqnarray*}

It turns out that, in the present model based on $f(R, T_\phi)$ theory, the qualitative signature-changing behavior is likewise controlled by the matrix $M$, in close analogy with the analysis of Dereli and Tucker \cite{Dereli}. In particular, when both eigenvalues of $M$ are positive, the corresponding $(X,Y)$-modes are purely hyperbolic and no regular signature transition is obtained. When the product of the eigenvalues is negative, the Hamiltonian constraint cannot be satisfied with real time-symmetric amplitudes, so this mixed sector is excluded from the regular signature-changing branch. By contrast, when both eigenvalues are negative, the minisuperspace variables typically display bounded oscillatory behaviour on one side of the transition and unbounded behaviour on the other, and this is precisely the regime in which one can choose parameters so that the metric is Euclidean for a finite interval, undergoes a signature change at the transition hypersurface, and then continues as Lorentzian for a finite interval thereafter, exactly in the spirit of the classification in \cite{Dereli}.

In the $\beta$-parametrization with the reconstructed variables $\tilde X(\beta):=X\!\big(t(\beta)\big),
\tilde Y(\beta):=Y\!\big(t(\beta)\big)$,
the scale factor and scalar field are written as
\begin{eqnarray}
\tilde a(\beta)&=&\Big(\tilde X(\beta)^2-\tilde Y(\beta)^2\Big)^{1/3},\nonumber\\
\tilde\phi(\beta)&=&\frac{1}{\sigma}\operatorname{arctanh}\!\left(\frac{\tilde Y(\beta)}{\tilde X(\beta)}\right).
\end{eqnarray}
Equivalently, defining $\tilde G(\beta):=\tilde X(\beta)^2-\tilde Y(\beta)^2$, 
we have $\tilde a(\beta)=\tilde G(\beta)^{1/3}$.

Following the Dereli--Tucker analysis, we now show that the signature-changing branch of our model is also geometrically meaningful in the later Kossowski--Kriele framework of transverse type-changing spacetimes \cite{Kriele}. Following the standard transverse notion of signature change, one requires that the metric be non-degenerate on each side of a hypersurface $\Sigma_0$ while $\det g$ vanishes on $\Sigma_0$ with first order so that the sign of $\det g$ flips across $\Sigma_0$, and that the radical (null) direction at $\Sigma_0$ be transverse to $\Sigma_0$. Indeed, in the $\beta$--parametrization of the metric \eqref{metric1}
with transition hypersurface
$\Sigma_0$, one has
\begin{equation}
\det g = (-\beta)\,\tilde a(\beta)^6\,\det\gamma .
\end{equation}
Hence, if $\tilde a(0)>0$, equivalently $\tilde G(0)>0$, then $\det g$ vanishes at $\Sigma_0$ with a simple zero and changes sign across $\beta=0$, so the metric undergoes a transverse change of signature. Moreover, at $\Sigma_0$ the radical is spanned by $\partial_\beta$, which is transverse to $T\Sigma_0$, since $\Sigma_0$ is given by $\beta=\mathrm{constant}$.
To examine regularity at the signature-change hypersurface, we have
\begin{equation}
\tilde a'(\beta)=\frac{1}{3}\,\tilde G(\beta)^{-2/3}\,\tilde G'(\beta).
\end{equation}
Now, by the chain rule and $t'(\beta)=\beta^{1/2}$, one finds $\tilde G'(\beta)=\frac{d}{d\beta}G\!\big(t(\beta)\big)
=\dot G\!\big(t(\beta)\big)\,t'(\beta)
=\beta^{1/2}\,\dot G\!\big(t(\beta)\big)$, 
where $G(t):=X(t)^2-Y(t)^2$. Therefore
\begin{equation}
\tilde a'(\beta)=\frac{1}{3}\,\tilde G(\beta)^{-2/3}\,\beta^{1/2}\,\dot G\!\big(t(\beta)\big).
\end{equation}
Assuming $\tilde G(0)>0$ and $\dot G(0)$ finite, it follows immediately that
\begin{equation}
\tilde a'(0)=0.
\end{equation}
Thus the induced spatial metric $h_{ij}(\beta)=\tilde a(\beta)^2\gamma_{ij}$ satisfies $
\partial_\beta h_{ij}\big|_{\Sigma_0}=0,$
and, equivalently, in cosmic time on the Lorentzian side, one has $\dot a|_{\Sigma_0}=0$. Hence the signature-change hypersurface $\Sigma_0$ is totally geodesic, i.e. $K_{ij}\big|_{\Sigma_0}=0$,
which is precisely the regularity condition required in the Kossowski--Kriele analysis to avoid distributional singular source terms at the transition surface \cite{Kriele}.

In the following, we show that an alternative choice of potential, an exponential potential, leads to novel solutions consistent with the conditions required for a signature change \cite{Kriele}. Exponential potentials play an important role in modern cosmology owing to both their firm theoretical origin and their remarkable dynamical properties. Such potentials arise generically in higher-dimensional and higher-curvature gravity theories, as well as in string-theoretic compactifications, where moduli fields frequently acquire effective exponential interactions due to internal curvature effects or non-perturbative phenomena such as gaugino condensation \cite{Whitt1984,Green1987,deCarlos1993}. From a cosmological standpoint, exponential potentials are particularly significant because they admit exact power-law inflationary solutions for sufficiently shallow slopes \cite{LucchinMatarrese1985,Halliwell1987}, while for steeper slopes they give rise to scaling (tracking) attractor solutions in which the scalar field energy density evolves proportionally to that of the dominant background fluid \cite{Wetterich1988,Copeland1998,Ferreira1997}. This attractor behavior renders the cosmological dynamics largely insensitive to initial conditions and has important implications for relic field abundances, nucleosynthesis constraints, and dark energy model building. Moreover, the associated dynamical system is autonomous and admits a complete phase-space characterization, allowing a rigorous analysis of late-time attractors and stability properties~\cite{Copeland1998}. Owing to their theoretical ubiquity and rich cosmological structure interpolating between inflationary and scaling regimes, exponential potentials therefore provide a well-motivated and robust framework for investigating both early and late universe dynamics.

Considering the Lagrangian \eqref{lag2}, one can introduce light--cone variables
\begin{equation}
u:=X+Y,\qquad v:=X-Y,
\qquad
X=\frac{u+v}{2},\quad Y=\frac{u-v}{2},
\qquad
X^2-Y^2=uv.
\label{eq:uvdef}
\end{equation}
Using $X=a^{3/2}\cosh(\sigma\phi)$ and $Y=a^{3/2}\sinh(\sigma\phi)$, one finds
\begin{equation}
u=a^{3/2}e^{\sigma\phi},\qquad v=a^{3/2}e^{-\sigma\phi},
\qquad
\frac{u}{v}=e^{2\sigma\phi}.
\label{eq:uvis}
\end{equation}

We now consider an exponential potential of the form
\begin{equation}
V(\phi)=V_0\,e^{2\gamma\sigma\phi},
\label{eq:Vgeneral}
\end{equation}
where $\gamma$ is either a positive or negative constant representing the growing and decaying branches. 
Since $e^{2\sigma\phi}=u/v$, the potential can be written as
\begin{equation}
V(\phi)=V_0\left(\frac{u}{v}\right)^{\gamma}.
\end{equation}
Hence, the minisuperspace potential term becomes
\begin{equation}
\frac{3B}{4}(X^2-Y^2)V(\phi)
=
\frac{3B}{4}\,uv\,V_0\left(\frac{u}{v}\right)^{\gamma}
=
\eta\,u^{1+\gamma}v^{1-\gamma},
\qquad
\eta:=\frac{3B}{4}V_0,
\label{eq:generaleta}
\end{equation}
and the kinetic part simplifies as 
\begin{equation}
\dot X^{\,2}-\dot Y^{\,2}
=\dot u\,\dot v.
\end{equation}
Therefore, the Lagrangian takes the compact form
\begin{equation}
\mathcal{L}(u,v,\dot u,\dot v)
=
\dot u\,\dot v
+
\eta\,u^{1+\gamma}v^{1-\gamma}.
\label{eq:Lgeneral}
\end{equation}
The Euler--Lagrange equations are then
\begin{equation}
\ddot v
=
\eta(1+\gamma)\,u^{\gamma}v^{1-\gamma},
\qquad
\ddot u
=
\eta(1-\gamma)\,u^{1+\gamma}v^{-\gamma}.
\label{eq:EOMgeneral}
\end{equation}
Here, without losing generality, one can consider two interesting cases $\gamma=\pm1$. For $\gamma=+1$, the field equations reduce to the system
\begin{equation}
\ddot u=0,
\qquad
\ddot v=2\eta\,u,
\end{equation}
with the solutions as
\begin{eqnarray}
u(t)&=&c_1 t+c_0,\nonumber\\
v(t)&=&\frac{\eta c_1}{3}t^3+\eta c_0 t^2+c_2 t+c_3,
\end{eqnarray}
where $c_0,c_1,c_2,c_3$ are integration constants. For $\gamma=-1$, the roles of $u$ and $v$ functions interchanges and the system takes the form
\begin{equation}
\ddot v=0,
\qquad
\ddot u=2\eta\,v,
\end{equation}
and consequently
\begin{eqnarray}
v(t)&=&b_1 t+b_0,\nonumber\\
u(t)&=&\frac{\eta b_1}{3}t^3+\eta b_0 t^2+b_2 t+b_3,
\end{eqnarray}
with $b_0,b_1,b_2$, and $b_3$ as integration constants. In this sense, the $\gamma=\pm1$ sectors are mapped into each other by the involution $u\leftrightarrow v$, and since $a^3=X^2-Y^2=uv$ the functional structure of the scale factor is the same in both cases, up to a relabelling of constants.
Imposing the Hamiltonian constraint $\mathcal{H}=\dot u\,\dot v-\eta\,u^2=0$ one obtains the constraint $c_1c_2=\eta\,c_0^{\,2}$. 
Thus, the field equations provide four integration constants and the Hamiltonian constraint
removes one, leaving three independent constants.

The corresponding solutions for the scale factor $a(t)$ and $\phi(t)$ will be \begin{eqnarray}
a(t) &=&\Big(X^2-Y^2\Big)^{\frac{1}{3}}=\Big(u(t)\,v(t)\Big)^{\frac{1}{3}}=
\left[
\frac{\eta c_1^2}{3} t^4
+ \frac{4\eta c_0 c_1}{3} t^3
+ 2\eta c_0^2 t^2
+ \left(c_1 c_3 + \frac{\eta c_0^3}{c_1}\right) t
+ c_0 c_3
\right]^{1/3},
\nonumber\\
\phi(t) &=& \frac{1}{2\sigma}\ln\!\left(\frac{u(t)}{v(t)}\right)=
\frac{1}{2\sigma}
\ln
\left(
\frac{c_1 t + c_0}
{
\frac{\eta c_1}{3}\, t^3
+ \eta c_0 t^2
+ \frac{\eta c_0^2}{c_1}\, t
+ c_3
}
\right),
\end{eqnarray}
where $c_1 \neq 0$.

Reality of $a$ and $\phi$ requires $u(t)v(t)\ge 0$ and $u(t),v(t)$ of the same sign on the interval of interest. In this scenario, the signature transition occurs at a finite size scale factor $a(0)=(c_0c_3)^\frac{1}{3}$ and for a nonzero scalar-field value $\phi(0)=\frac{1}{2\sigma}\ln\!\left(\frac{c_0}{c_3}\right)$. 

Using the transformation $t(\beta)=\frac{2}{3}\,\beta^{3/2}$, we also find
\begin{eqnarray}
\tilde a(\beta)&=&\left[
\frac{16\eta c_1^2}{243}\,\beta^6
+\frac{32\eta c_0 c_1}{81}\,\beta^{9/2}
+\frac{8\eta c_0^2}{9}\,\beta^3
+\frac{2}{3}\left(c_1 c_3+\frac{\eta c_0^3}{c_1}\right)\beta^{3/2}
+c_0 c_3
\right]^{1/3},\nonumber\\
\tilde \phi(\beta)&=&\frac{1}{2\sigma}\ln\!\left(
\frac{\frac{2c_1}{3}\beta^{3/2}+c_0}
{
\frac{8\eta c_1}{81}\beta^{9/2}
+\frac{4\eta c_0}{9}\beta^3
+\frac{2\eta c_0^2}{3c_1}\beta^{3/2}
+c_3
}
\right).
\end{eqnarray}

Similar to the previous solutions, here one can show that the solutions for the exponential potential are also compatible with Kossowski--Kriele conditions for  transverse type-changing spacetimes \cite{Kriele}. Remember the metric \eqref{metric1}, and its determinant 
\begin{equation*}
\det g =\big(-\beta\big)\,\tilde a(\beta)^6\,\det\gamma.
\end{equation*}
Consider the Lorentzian region parametrized by $\beta\ge 0$ with $t(\beta)=\frac{2}{3}\,\beta^{3/2}$,
and define the scale factor by
\begin{equation}
a(t)=\tilde a(\beta(t)),
\qquad 
\tilde a(\beta)=\big(\tilde F(\beta)\big)^{1/3},
\qquad
\tilde F(\beta):=\tilde u(\beta)\,\tilde v(\beta),
\end{equation}
where 
\begin{align}
\tilde u(\beta)&:=u\!\big(t(\beta)\big)
=\frac{2c_1}{3}\,\beta^{3/2}+c_0,\\[1mm]
\tilde v(\beta)&:=v\!\big(t(\beta)\big)
=\frac{8\eta c_1}{81}\,\beta^{9/2}
+\frac{4\eta c_0}{9}\,\beta^{3}
+\frac{2c_2}{3}\,\beta^{3/2}
+c_3.
\end{align}
One observes that if $\tilde a(0)>0$, equivalently $\tilde F(0)=\tilde u(0)\tilde v(0)=c_0c_3>0$ , then $\det g\neq 0$ for $\beta\neq 0$ near $\Sigma_0$, while $\det g=0$ at $\beta=0$ with a simple zero and $\mathrm{sign}(\det g)$ changes with $\beta$, implying a change between Euclidean ($\beta<0$) and Lorentzian ($\beta>0$) signature.
Moreover, similar to the quadratic potential case, at $\Sigma_0$ the radical is spanned by $\partial_\beta$ since $g_{\beta\beta}\!\mid_{\Sigma_0}=0$ while the spatial block remains positive definite; because $\Sigma$ is given by $\beta=\mathrm{constant}$, $\partial_\beta$ is not tangent to $\Sigma_0$, hence the radical intersects $T\Sigma_0$ transversely.
Therefore the metric defined by our solution realizes a transverse signature change across $\Sigma_0$, with a well-defined Lorentzian domain $\beta>0$ provided $\tilde F(\beta)\ge 0$ there. Moreover, since $\tilde a'(\beta)\big|_{\Sigma_0}=0$, one has $K_{ij}\big|_{\Sigma_0}=0$, i.e. the hypersurface $\Sigma$ is totally geodesic. Hence the field equations do not generate singular source terms at $\Sigma_0$, and the energy--momentum tensor remains bounded across the signature-change surface according to \cite{Kriele}.
\\

Finally, we present the following no-go Lorentzian bounce theorem for this model.
\\
\textbf{Theorem:}  There exists no $\beta_b>0$ such that $\tilde a$ has a non-singular cosmological bounce at $\beta_b$, i.e.
\begin{equation}
\tilde a'(\beta_b)=0,\qquad \tilde a''(\beta_b)>0,\qquad \tilde F(\beta_b)>0.
\end{equation}
Moreover, a non-singular bounce cannot occur at the type-changing hypersurface $\beta=0$ either. 
\\
The proof of the theorem is given in the Appendix. 
The theorem shows that, within the exponential-potential branch subject to the constraint $c_1c_2=\eta c_0^2$, the signature-change construction does not provide a non-singular Lorentzian bounce. In the Lorentzian region ($\beta>0$), any candidate extremum of the scale factor necessarily satisfies $\tilde F'(\beta_b)=0$, but then $\tilde F(\beta_b)$ and $\tilde F''(\beta_b)$ acquire fixed signs controlled by $\eta$: for $\eta>0$ one finds $\tilde F(\beta_b)<0$, so the scale factor becomes non-real and the extremum is unphysical, whereas for $\eta<0$ one has $\tilde F(\beta_b)>0$ but $\tilde F''(\beta_b)<0$, implying that the extremum is a local maximum rather than a bounce. At the signature-change hypersurface $\beta=0$, the intrinsic degeneracy of the parametrization and the regularity requirement further obstruct a bounce interpretation. Consequently, this model class can realize a smooth signature transition, but it cannot mediate a contracting to expanding transition in the Lorentzian domain.

\section{Conclusion}
In this work, we investigated classical signature change in the trace-coupled modified gravity model $f(R,T_\phi)=R+\alpha T_\phi$ with a scalar-field source in an FLRW minisuperspace. Using the type-changing FLRW metric written in terms of the parameter $\beta$, where $\beta<0$ corresponds to the Euclidean region, $\beta>0$ to the Lorentzian region, and $\beta=0$ defines the signature-change hypersurface, we showed that the model admits real solutions that extend smoothly across this transition surface. A key step in the analysis was the hyperbolic field redefinition from the original cosmological variables, namely the scale factor $a$ and the scalar field $\phi$, to the new minisuperspace variables $X=a^{3/2}\cosh(\sigma\phi)$ and $Y=a^{3/2}\sinh(\sigma\phi)$, which recasts the system into an effective two-dimensional minisuperspace model. Within this formulation, we found that the trace-coupling parameter $\alpha$ does not merely deform the equations quantitatively, but also introduces qualitatively distinct regimes, especially at the critical values $\alpha=-\tfrac12$ and $\alpha=-\tfrac14$, where the scalar dynamics changes in a singular way.

We first considered a quadratic potential, and showed that the existence of regular signature-changing solutions is controlled by the eigenvalue structure of the effective minisuperspace matrix, in close analogy with the Dereli--Tucker analysis. In particular, the admissible branch arises when the parameters are such that the solutions display the bounded and unbounded behavior required for a finite Euclidean region followed by a Lorentzian phase. We also showed that these solutions are geometrically regular in the Kossowski--Kriele sense: the determinant of the metric vanishes linearly at the transition surface, the radical direction is transverse, and the hypersurface is totally geodesic. This ensures that the signature change is not a merely formal feature of the reduced variables, but corresponds to a well-defined and regular type change in spacetime geometry.

We further considered an exponential potential and, after introducing light-cone variables, obtained exact closed-form solutions in the integrable cases. These solutions again provide explicit examples of smooth Euclidean--Lorentzian transition in the same trace-coupled framework. However, we showed that this branch cannot support a non-singular Lorentzian bounce: although signature change remains possible, the structure of the exact solutions prevents the existence of a regular bouncing phase. Altogether, our results show that $f(R,T_\phi)$ gravity preserves the possibility of regular classical signature change beyond the Einstein-scalar case, while at the same time introducing new dynamical restrictions and critical structures that arise directly from matter-geometry coupling.

\appendix
\section*{Appendix}

\textbf{Proof of the Theorem:}
For $\beta>0$, differentiating $\tilde a(\beta)=(\tilde F(\beta))^{1/3}$ yields
\begin{equation*}
\tilde a'(\beta)=\frac{1}{3}\,\tilde F(\beta)^{-2/3}\tilde F'(\beta),
\qquad
\tilde a''(\beta)
=\frac{1}{3}\,\tilde F(\beta)^{-2/3}\tilde F''(\beta)
-\frac{2}{9}\,\tilde F(\beta)^{-5/3}\big(\tilde F'(\beta)\big)^2.
\end{equation*}
Hence, at any candidate $\beta_b>0$ with $\tilde a'(\beta_b)=0$, equivalently $\tilde F'(\beta_b)=0$, and $\tilde F(\beta_b)>0$, one has
\begin{equation*}
\tilde a''(\beta_b)=\frac{1}{3}\,\tilde F(\beta_b)^{-2/3}\tilde F''(\beta_b),
\end{equation*}
so a non-singular bounce in the Lorentzian region is equivalent to
\begin{equation*}
\tilde F(\beta_b)>0,\qquad \tilde F'(\beta_b)=0,\qquad \tilde F''(\beta_b)>0,
\qquad (\beta_b>0).
\end{equation*}

For the exponential-potential branch,
\begin{equation*}
\tilde u(\beta)=c_0+\frac{2c_1}{3}\,\beta^{3/2},
\qquad
\tilde v(\beta)=\frac{8\eta c_1}{81}\,\beta^{9/2}
+\frac{4\eta c_0}{9}\,\beta^{3}
+\frac{2c_2}{3}\,\beta^{3/2}
+c_3,
\qquad
c_1c_2=\eta c_0^2.
\end{equation*}
Differentiating $\tilde u$ and $\tilde v$,
\begin{equation*}
\tilde u'(\beta)=c_1\beta^{1/2},
\qquad
\tilde v'(\beta)=\frac{4\eta c_1}{9}\,\beta^{7/2}
+\frac{4\eta c_0}{3}\,\beta^{2}
+c_2\beta^{1/2}.
\end{equation*}
Using the constraint $c_2=\eta c_0^2/c_1$ and $\tilde u(\beta)=c_0+\frac{2c_1}{3}\beta^{3/2}$, one finds the exact identity
\begin{equation*}
\tilde v'(\beta)=\frac{\eta}{c_1}\,\tilde u(\beta)^2\,\beta^{1/2}.
\end{equation*}
Therefore,
\begin{equation*}
\tilde F'(\beta)=(\tilde u\tilde v)'
=\tilde u'(\beta)\tilde v(\beta)+\tilde u(\beta)\tilde v'(\beta)
=c_1\beta^{1/2}\,\tilde v(\beta)+\frac{\eta}{c_1}\,\tilde u(\beta)^3\,\beta^{1/2}.
\end{equation*}
At any Lorentzian extremum $\beta_b>0$ satisfying $\tilde F'(\beta_b)=0$, we can divide by $\beta_b^{1/2}\neq 0$ to obtain
\begin{equation*}
c_1\tilde v(\beta_b)=-\frac{\eta}{c_1}\,\tilde u(\beta_b)^3,
\qquad\Longrightarrow\qquad
\tilde F(\beta_b)=\tilde u(\beta_b)\tilde v(\beta_b)
=-\frac{\eta}{c_1^2}\,\tilde u(\beta_b)^4.
\end{equation*}
Substituting $\tilde u(\beta)=c_0+\frac{2c_1}{3}\beta^{3/2}$ yields the purely-$\beta$ form
\begin{equation}
\tilde F'(\beta_b)=0
\ \Longrightarrow\
\tilde F(\beta_b)
=
-\frac{\eta}{c_1^2}\left(c_0+\frac{2c_1}{3}\beta_b^{3/2}\right)^4,
\qquad (\beta_b>0).
\end{equation}
In particular, at any Lorentzian extremum $\beta_b>0$,
\begin{equation}
\mathrm{sign}\!\big(\tilde F(\beta_b)\big)=-\mathrm{sign}(\eta).
\end{equation}
In particular
\begin{itemize}
\item If $\eta>0$, then $\tilde F(\beta_b)<0$ at any extremum, which violates the Lorentzian reality requirement $\tilde F(\beta_b)\ge 0$.
\item If $\eta<0$, then $\tilde F(\beta_b)>0$ at any extremum, but (as shown next) the extremum is necessarily a local maximum rather than a bounce.
\end{itemize}

Direct differentiation gives
\begin{equation*}
F''(t)=4\eta\,u(t)^2.
\end{equation*}
Using the chain rule for $\tilde F(\beta)=F(t(\beta))$ with $dt/d\beta=\beta^{1/2}$ and $d^2t/d\beta^2=\frac12\beta^{-1/2}$, one finds for $\beta>0$
\begin{equation*}
\tilde F''(\beta)=\beta\,F''(t(\beta))+\frac{1}{2}\beta^{-1/2}F'(t(\beta)).
\end{equation*}
Thus, at any extremum $\beta_b>0$, equivalently $F'(t_b)=0$, we obtain
\begin{equation*}
\tilde F''(\beta_b)=\beta_b\,F''(t_b)=4\eta\,\beta_b\,\big(\tilde u(\beta_b)\big)^2.
\end{equation*}
Hence $\tilde F''(\beta_b)$ has the same sign as $\eta$ for $\beta_b>0$. Therefore, one can conclude the result as
\begin{itemize}
\item If $\eta>0$, then $\tilde F''(\beta_b)>0$ but $\tilde F(\beta_b)<0$ at any extremum, so a bounce is impossible because the scale factor becomes non-real.
\item If $\eta<0$, then $\tilde F(\beta_b)>0$ at any extremum, but $\tilde F''(\beta_b)<0$, so the extremum is a local maximum, not a bounce.
\end{itemize}
Therefore, no $\beta_b>0$ can satisfy simultaneously $\tilde F(\beta_b)>0$, $\tilde F'(\beta_b)=0$, and $\tilde F''(\beta_b)>0$. This proves the Lorentzian no-go.

At $\beta=0$ one has $t(0)=0$ and $\tilde F(0)=F(0)=u(0)v(0)=c_0c_3$.
Moreover,
\begin{equation*}
\tilde F'(\beta)=\beta^{1/2}\dot F(t(\beta)),
\qquad
\tilde F''(\beta)=\beta\,\ddot F(t(\beta))+\frac{1}{2}\beta^{-1/2}\dot F(t(\beta)).
\end{equation*}
Hence $\tilde F''$ is generically singular as $\beta\to 0^+$ unless $\dot F(0)=0$.
Using $c_1c_2=\eta c_0^2$,
\begin{equation*}
\dot F(0)=c_1c_3+c_0c_2
=c_1c_3+\frac{\eta c_0^3}{c_1},
\end{equation*}
so regularity at $\beta=0$ requires $c_1^2c_3+\eta c_0^3=0$, i.e.
\begin{equation}
c_3=-\frac{\eta c_0^3}{c_1^2}
\qquad\Longrightarrow\qquad
F(0)=c_0c_3=-\frac{\eta}{c_1^2}\,c_0^4.
\end{equation}
Thus, if $\eta>0$ then $F(0)<0$ under the regularity condition and the scale factor is not real at $\beta=0$; whereas if $\eta<0$ then $F(0)>0$ but $\ddot F(0)=4\eta c_0^2<0$ so the extremum corresponds to a local maximum rather than a bounce. In any case, $\beta=0$ is intrinsically degenerate and does not support a regular Lorentzian bounce.

\newcommand{\bibTitle}[1]{``#1''}
\begingroup
\let\itshape\upshape
\bibliographystyle{plain}

\end{document}